\documentclass[prb,twocolumn,amsmath,amssymb,showpacs]{revtex4}

\usepackage{graphicx}
\usepackage{dcolumn}
\usepackage{bm}
\usepackage{mathrsfs}
\usepackage{appendix}
\usepackage{bbm}
\usepackage{color}

\def \vo {{\bf o}}

\def \vH {{\bf B}}

\def \vS {{\bf S}}

\def \vQ {{\bf Q}}

\def \vq {{\bf q}}

\def \vm {{\bf m}}
\def \vo {{\bf o}}

\def \mb {\mu_{\rm B}}

\def \vB {{\bf B}}
\def \vE {{\bf E}}
\def \Dkp {D^{\rm kag}}
\def \Dtp {D^{\rm tri}}
\def \Dka {C^{\rm kag}}
\def \Dth {A^{\rm tri}}
\def \Dta {C^{\rm tri}}

\def \va {{\bf a}}
\def \vb {{\bf b}}
\def \vc {{\bf c}}

\def \vM {{\bf M}}
\def \vP {{\bf P}}

\def \CBO {{\rm CaBaCo$_4$O$_7$} }
\def \CBP {{\rm CaBaCo$_4$O$_7$}}

\def \vE {{\bf E}}

\def \vn {{\bf n}}
\def \vz {{\bf z}}
\def \vx {{\bf x}}
\def \vy {{\bf y}}

\def \mB {\mu_{\rm B}}

\begin{document}

\title{{Competing Exchange Interactions in the Multiferroic and Ferrimagnetic CaBaCo$_4$O$_7$}\footnote{Copyright notice: This manuscript has been authored by UT-Battelle, LLC under Contract No. DE-AC05-00OR22725 with the U.S. Department of Energy. The United States Government retains and the publisher, by accepting the article for publication, acknowledges that the United States Government retains a non-exclusive, paid-up, irrevocable, world-wide license to publish or reproduce the published form of this manuscript, or allow others to do so, for United States Government purposes. The Department of Energy will provide public access to these results of federally sponsored research in accordance with the DOE Public Access Plan (http://energy.gov/downloads/doe-public-access-plan).}}

\author{R.S. Fishman$^1$, S. Bord\'acs$^2$, V. Kocsis$^2$, I. K\'ezsm\'arki$^2$,  J. Viirok$^3$, U. Nagel$^3$, T. R\~o\~om$^3$, A. Puri$^4$, U. Zeitler$^4$,
Y. Tokunaga$^{5,6}$, Y. Taguchi$^5$,  and Y. Tokura$^{5,7}$
}

\affiliation{$^1$Materials Science and Technology Division, Oak Ridge National Laboratory, Oak Ridge, Tennessee 37831, USA}
\affiliation{$^2$Department of Physics, Budapest University of Technology and Economics and MTA-BME Lend\"ulet Magneto-optical Spectroscopy
Research Group, 1111 Budapest, Hungary}
\affiliation{$^3$National Institute of Chemical Physics and Biophysics, Akadeemia tee 23, 12618 Tallinn, Estonia}
\affiliation{$^4$High Field Magnet Laboratory (HFML-EMFL), Radboud University Nijmegen, Toernooiveld 7, 6525 ED Nijmegen, The Netherlands}
\affiliation{$^5$RIKEN Center for Emergent Matter Science (CEMS), Wako, Saitama 351-0198, Japan}
\affiliation{$^6$Department of Advanced Materials Science, University of Tokyo, Kashiwa 277-8561, Japan}
\affiliation{$^7$Department of Applied Physics, University of Tokyo, Hongo, Tokyo 113-8656, Japan}

\date{\today}

\begin{abstract}
Competing exchange interactions can produce complex magnetic states together with spin-induced electric polarizations.  
With competing interactions on alternating triangular and kagome layers, the swedenborgite \CBO may have one of the largest measured 
spin-induced polarizations of $\sim 1700$ nC/cm$^2$ below its ferrimagnetic transition temperature at 70 K.  Upon rotating our sample about $c=[0,0,1]$ 
while the magnetic field is fixed along $[1,0,0]$,
the three-fold splitting of the spin-wave frequencies indicates that our sample is hexagonally twinned.  Magnetization measurements then indicate that 
roughly 20\% of the sample is in a domain with the $a$ axis along $[1,0,0]$ and that 80\% of the sample is in one of two other domains with the $a$ axis along either $[-1/2,\sqrt{3}/2,0]$
or $[-1/2,-\sqrt{3}/2,0]$.  Powder neutron-diffraction data, magnetization measurements, and THz absorption spectroscopy reveal that the 
complex spin order in each domain can be described as a triangular array of bitetrahedral $c$-axis chains ferrimagnetically coupled to each other in the $ab$ plane. 
The electric-field dependence of bonds coupling those chains produces the large spin-induced polarization of \CBP.  

\end{abstract}


\maketitle

\section{Introduction}

Competing exchange interactions produce complex magnetic states with a wide range of interesting behavior found in 
spin glass [\onlinecite{binder86}], spin ice [\onlinecite{harris97}], and magnetic skyrmions [\onlinecite{muhl11}].  In multiferroic materials, 
complex spin states can exibit a spin-induced electric polarization $\vP $ due to either
the spin current, $p$-$d$ orbital hybridization, or magnetostriction [\onlinecite{khomskii06, cheong07}].  Because the coupling between the 
electrical and magnetic properties in multiferroic materials is both scientifically and technologically important, the effects of 
competing exchange interactions have been investigated in a wide range of multiferroic materials such as 
$R$MnO$_3$ [\onlinecite{sushkov07}], CoCrO$_4$ [\onlinecite{yama06}], CuCrO$_2$ [\onlinecite{soda10}], CuFeO$_2$ [\onlinecite{seki10}],
and MnWO$_4$ [\onlinecite{taniguchi09}].  While the first four materials [\onlinecite{sushkov07, yama06, soda10, seki10}] are geometrically frustrated due to 
competing interactions on a triangular lattice, MnWO$_4$ [\onlinecite{taniguchi09}] exhibits long-range competing interactions [\onlinecite{ye2011}] 
on a highly-distorted monoclinic lattice.

\begin{figure}
\includegraphics[width=8cm]{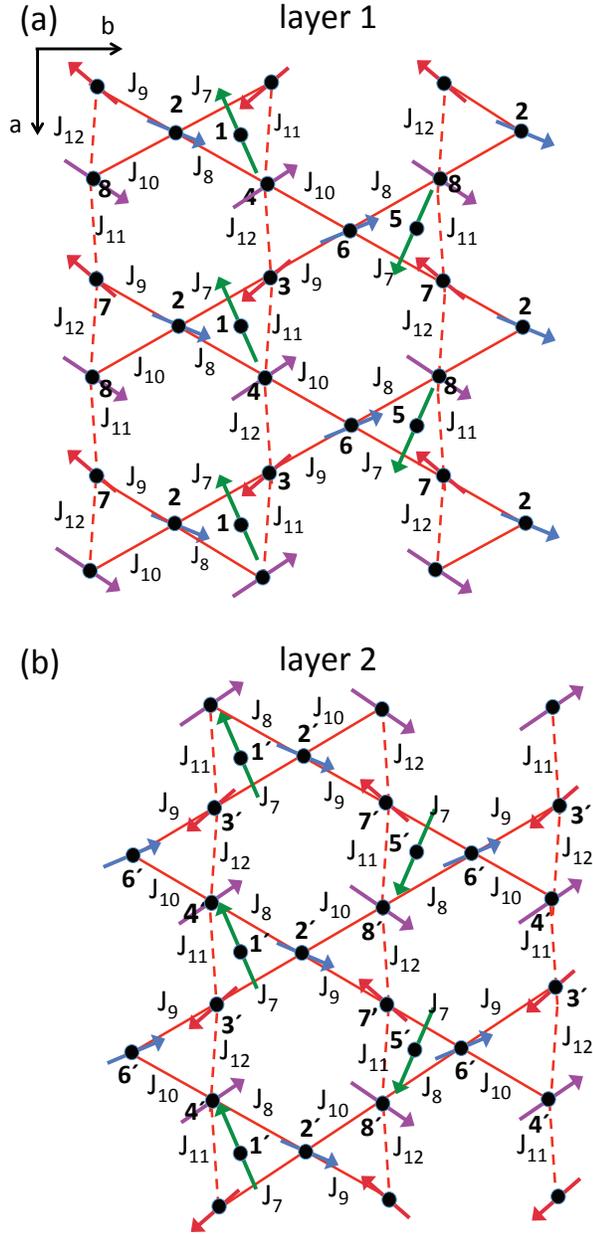}
\caption{(Color online) (a) and (b) The predicted spin configuration for layers 1 and 2 in zero field.   Spins 1 and 5 lie on a triangular layer
above the first kagome layer in (a);  spins 1$^{\prime }$ and 5$^{\prime }$ lie on a triangular layer above the second kagome layer in (b).
Layers are arranged so that spins 1$^{\prime }$ and 5$^{\prime }$ lie directly above spins 1 and 5.
}
\end{figure}

Compounds in the ``114" swedenborgite family [\onlinecite{rav12}] $R$Ba$M$$_4$O$_7$ ($M$= Co or Fe) 
contain alternating triangular and kagome layers, 
both of which are geometrically frustrated when undistorted.  
The ``114" cobaltites [\onlinecite{val02, tsipis05, bych05}] 
were initially studied to find charge ordering among the Co$^{2+}$ and Co$^{3+}$ ions.  An important member of this family,
YBaCo$_4$O$_7$ exhibits antiferromagnetic ordering [\onlinecite{chap06, khal11}] below 110 K and diffuse scattering [\onlinecite{val02, tsipis05}] 
indicative of spin disorder below 60 K.  
The magnetic state between 110 K and 60 K is stabilized by a structural transition [\onlinecite{bera13}] that relieves the geometric frustration.  
Both structural and magnetic transitions are quite sensitive to excess oxygen and no magnetic order [\onlinecite{maig06, avci13}] appears in 
YBaCo$_4$O$_{7+\delta }$ for $\delta \ge 0.12$.  Another family member, YbBaCo$_4$O$_7$ undergoes a structural transition at 175 K 
that stabilizes an antiferromagnetic state below 80 K  [\onlinecite{huq06}].

A particularly interesting ``114" cobaltite, \CBO undergoes an orthorhombic distortion [\onlinecite{caig10, panja16}] that
relieves the geometric magnetic frustration on both the  
kagome and triangular layers sketched in Fig.~1 above the magnetic transition temperature 
$T_c = 70$ K.  Below $T_c$, \CBO develops a very large spin-induced 
polarization $\sim 1700$ nC/cm$^2$ [\onlinecite{caig13}], second only to the conjectured  [\onlinecite{jun15}] spin-induced polarization $\sim 3000$ nC/cm$^2$ of BiFeO$_3$.  
 Also unusual, \CBO displays a substantial ferrimagnetic moment of about 0.9 $\mb $ per formula unit (f.u.) [\onlinecite{caig09}], 
which could allow magnetic control of the electric polarization.
Although its ferroelectric transition is inaccessible and its permanent electric polarization is not switchable [\onlinecite{john14}],
applications of \CBO might utilize the large spin-induced polarization produced by a magnetic field just below $T_c$ [\onlinecite{caig13}].

This paper examines the magnetic properties of \CBO based on a Heisenberg model with 12 nearest neighbor interactions and 
associated anisotropies.  The magnetic state of \CBO can be described as a triangular array of ferrimagnetically aligned, bitetrahedral $c$-axis 
chains with net moment along $\vb $.  Competing interactions within each chain produce a non-collinear spin state.
The strong electric polarization of \CBO below $T_c$ is induced by the displacement of oxygen atoms surrounding bonds that couple those chains. 

This paper has six sections.
Section II proposes a microscopic model for \CBP.  New magnetization and optical measurements are 
presented in Section III.   Fitting results are discussed in Section IV.  In Section V, we predict the spin-induced electric polarization.  Section VI contains a
conclusion.

\section{Microscopic Model}

Each magnetic unit cell of \CBO contains 16 Co ions on two kagome and two triangular layers with orthorhombic 
lattice constants $a=6.3$ \AA , $b=11.0$ \AA , and $c=10.2$ \AA .  Four crystallographically distinct Co ions have three 
different valences [\onlinecite{caig10, chat11}].  Triangular layers contain mixed-valent Co$^{3+}$/Co$^{2+}\underline{L}$ (${\underline L}$ is a ligand hole) spins 1, 5, 9, and 13 
with moments $M_1=2.9\,\mb $.  Kagome layers contain Co$^{2+}$ spins 2, 3, 6, 7, 10, 11, 14, and 15 with moments $M_2=M_3=2\,\mb $ 
and mixed-valent Co$^{3+}$/Co$^{2+}\underline{L}$ spins 4, 8, 12, and 16 with $M_4=2.4\,\mb $.   
Because adjacent kagome or triangular layers are related by symmetry, $\vS_{i^{\prime }}=\vS_{i+8}$ on layer two is identical to $\vS_i$ on layer one.
With $\vS_i = S_i (\cos \phi_i ,\sin \phi_i , 0)$ constrained to the $ab$ plane, the ferrimagnetic moment lies along $\vb$ if
$\phi_{i+4} = \pi -\phi_i$ ($i=1,\ldots, 4$).

The 12 different nearest-neighbor exchange couplings $J_i$ are drawn in Figs.~1(a-b) and 2.  Six of these ($J_1$ through $J_6$) couple the kagome and triangular layers as shown in
Fig.~2;  the other six ($J_7$ through $J_{12}$) couple the spins within a kagome layer as shown in Figs.~1(a) and (b).  
The dominance of nearest-neighbor exchange over next-nearest neighbor exchange [\onlinecite{john14}]
justifies setting the exchange interactions between spins on the triangular layers to zero.
Our model also includes easy-plane anisotropies $D$, easy-axis 
anisotropies $C$ within both kagome and triangular layers, and hexagonal anisotropy $A$ on the triangular layers.

With magnetic field $\vH $ along $\vm $, the Hamiltonian is
\begin{eqnarray}
{\cal H}&=&-\sum_{\langle i,j\rangle }J_{ij} \vS_i \cdot \vS_j  +\Dtp \sum_{i, {\rm tri}} {S_{ic}}^2 + \Dkp \sum_{i, {\rm kag}} {S_{ic}}^2 \nonumber \\
&-& \Dka \sum_{i, {\rm kag}} (\vo_i \cdot \vS_i )^2 
-\Dta \sum_{i, {\rm tri}} (\vn_i \cdot \vS_i )^2  \nonumber \\
&-& \Dth \,{\rm{Re}} \sum_{i, {\rm tri}} \bigl(S_{ia}+iS_{ib}\bigr)^6 - g\mb B \sum_i  \vm \cdot \vS_i,
\end{eqnarray}
where $\vS_i$ is a spin $S$ operator on site $i$.  For simplicity, we set $g=2$ for all spins.

\begin{figure}
\includegraphics[width=8cm]{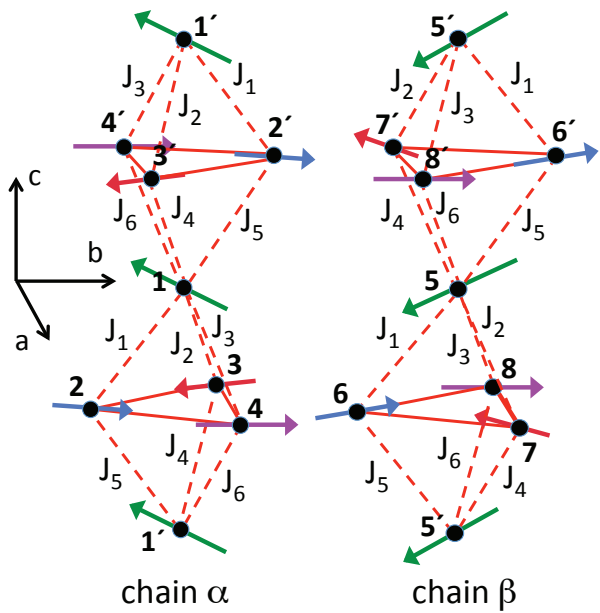}
\caption{(Color online) A sideways view of the zero-field spin configuration showing bitetrahedral $c$-axis chains $\alpha $ and $\beta $.  
}
\end{figure}

The easy-axis anisotropy terms proportional to $\Dka $ and $\Dta $ involve unit vectors
$\vo_i$ along the ``bowtie" directions $\phi_i = \pi /2 $ (spins 2 and 6), $5\pi /6$ (spins 3 and 8) and $7\pi /6$ (spins 4 and 7) for the kagome layers
and $\vn_i$ along the $\phi_i= \pi/6$ (spin 1) and $-\pi/6 $ (spin 5) directions for the triangular layers.  The hexagonal anisotropy 
on the triangular layers has expectation value 
\begin{equation}
-\Dth {S_1}^6 \sum_{i ,{\rm tri}} \sin^6 \theta_i \cos 6\phi_i .\nonumber 
\end{equation}
All anisotropy terms may act to constrain the spins to the $ab$ plane.  

Spin amplitudes $S_n$ are fixed at their observed values $M_n/2\mb $ after performing a $1/S$ expansion about the classical limit.  Alternatively,
the spins $S_n$ could all have been taken as 3/2 but with different $g$-factors for different sets of spins.  As discussed below, that would reduce
the estimated exchange coupling $J_{ij}$ by a factor of $4S_iS_j/9$.

Static properties are obtained by minimizing the classical energy $\langle {\cal H}\rangle $ (the zeroth-order term in this expansion) with respect to the 16 spin angles.
The eigenvalues and eigenvectors of a $32 \times 32$ equations-of-motion matrix [\onlinecite{fishman13}] produced by the second-order term in the $1/S$ expansion
give the optical mode frequencies and absorptions, respectively.

\section{Magnetization and optical measurements}

Perhaps due to excess or deficient oxygen [\onlinecite{seikh14}] or different domain populations (see below), 
previous magnetization measurements [\onlinecite{caig09, caig10, qu11, iwamoto12, seikh12}] on \CBO 
are rather scattered.  Consequently, new magnetization measurements were performed at 4 K on hexagonally-twinned crystals with a common $\vc =\vz  =[0,0,1]$ axis.  
In domain I, $\va $ lies along the laboratory direction $\vx = [1,0,0]$ and $\vb $ lies along $\vy = [0,1,0]$;  in domain II, $\va = [-1/2,\sqrt{3}/2, 0]$ and $\vb = [-\sqrt{3}/2,-1/2,0]$;
and in domain III, $\va = [-1/2,-\sqrt{3}/2, 0]$ and $\vb = [\sqrt{3}/2,-1/2,0]$.  
If $p_l$ are the domain populations, then the magnetizations $M_x$ and $M_y$ measured with fields along $\vx $ and $\vy $
only depend on $p_1$ and $p_2 + p_3 = 1-p_1$.  Of course, $M_z$ measured with field along $\vz $ is independent of
$p_l$.  Fig.~3 indicates that all three magnetizations increase monotonically up to at least 32 T.  

\begin{figure}
\includegraphics[width=8.8cm]{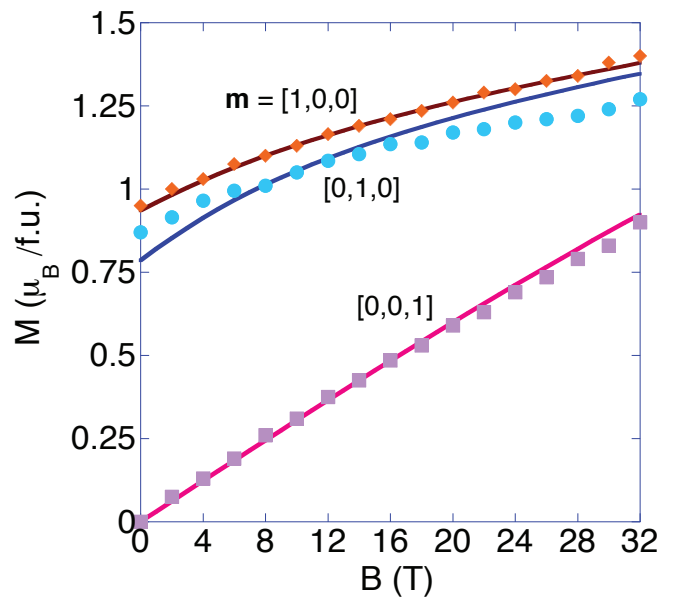}
\caption{(Color online) 
The measured (circles and squares) and predicted (solid curves) magnetizations for field along $[1,0,0]$, $[0,1,0]$, or $[0,0,1]$.
}
\end{figure}

Previous optical measurements [\onlinecite{bord15}] at the ordering wavevector $\vQ $ found two
conventional spin-wave modes that couple to the ground state through the magnetization operator $\vM =2\mB \sum_i \vS_i$.
These magnetic-resonance (MR) modes are degenerate in zero field with a frequency of 1.07 THz and split almost linearly with 
increasing field along $\vy $, as shown in Fig.~4.  
For $\vm= \vy $, the MR modes are excited in two geometries:  ($i$) with THz fields $\vE_{\omega } || \vx $ 
and $\vB_{\omega } || \vz $ and ($ii$) with $\vE_{\omega } || \vz $ and $\vB_{\omega } || \vx $.  Those measurements also found 
an electromagnon (EM) that couples to the ground state through the polarization operator $\vP $.  The EM with zero-field frequency 1.41 
THz is only excited in geometry $ii$.   

\begin{table*}
\caption{Exchange and anisotropy parameters in meV.}
\begin{ruledtabular}
\begin{tabular}{ccccccccccccccc}
& $p_1$ & $J_1=J_5$ & $J_2=J_4$ & $J_3=J_6$  & $J_7$ & $J_8$ & $J_9$ & $J_{10}$ & $J_{11}=J_{12}$ & $\Dkp $ & $\Dtp $ & $\Dka $ & $\Dta $ & ${S_1}^4\Dth $   
  \\
 \hline                            
 & 0.185 & $-91.5$  & $-10.8$ & $41.4$  & $-29.9$ & $187.8$ & $7.9$ & $108.0$ & $-6.7$ & $-0.67$ & $-1.24$ & $3.70$ & $0.77$ & $0.0064$ \\
 error & $\pm 0.071 $ & $\pm 3.6$ & $\pm 0.5$ & $\pm 0.7$  &  $\pm 1.9$ & $\pm 7.5$ & $\pm 0.1$ & $\pm 4.5$ &  $\pm 0.1$ & $\pm 0.06$ & $\pm 0.03$ & $\pm 0.15$ & $\pm 0.09$ & $\pm 0.0004$
\\
\end{tabular}
\end{ruledtabular}
\label{all}
\end{table*}

\begin{figure}
\includegraphics[width=8cm]{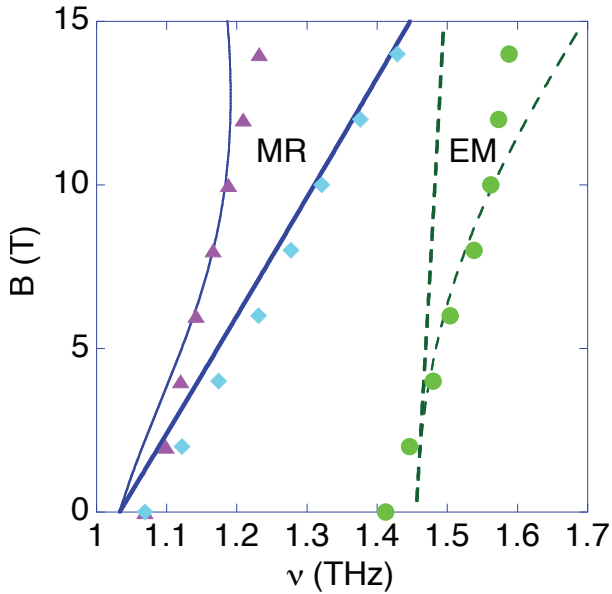}
\caption{(Color online) The predicted MR (solid) and EM (dashed) modes for domain I (thick) and domains II and III (thin).
Measured modes are indicated by solid points.
}
\end{figure}

Because the exchange couplings already break every degeneracy in the unit cell,
the 16 predicted modes for a single domain are non-degenerate.  Therefore, the split MR modes must come from different domains.  
This was verified by measuring [\onlinecite{experiment}] the MR mode frequencies as a function of the rotation angle $\theta $ for field 
$\vB =B(\cos \theta ,\sin \theta ,0)=B(\vx \cos \theta + \vy \sin \theta )$ in the laboratory reference frame.  In practice, this is accomplished
by rotating the sample about $\vc $ while keeping the field fixed along $\vx $.
As shown in Fig.~5 for 12 and 15 T, each hexagonal domain then contributes one MR branch with a period of $\pi $.  

With field $\vB_{{\rm loc}} = B (\cos \psi , \sin \psi, 0)= B(\va \cos \psi + \vb \sin \psi )$ in the domain reference frame, 
the upper MR mode in Fig.~4 corresponds to the $\psi = \pi /2$ mode 
for domain I while the lower MR mode corresponds to the degenerate $\psi = \pm \pi /6$ modes for domains II and III.    
Previously measured MR frequencies plotted in Fig.~4 at 12 T correspond to the 
diamond and triangular points in Fig.~5(b) at $\theta = \pi/2$. 
Cusps in the MR curves for each domain at $\psi = 0$ and $\pi $ are caused by flipping the $b$ component of the magnetization (see inset to Fig.~5(b)).

\begin{figure}
\includegraphics[width=8.3cm]{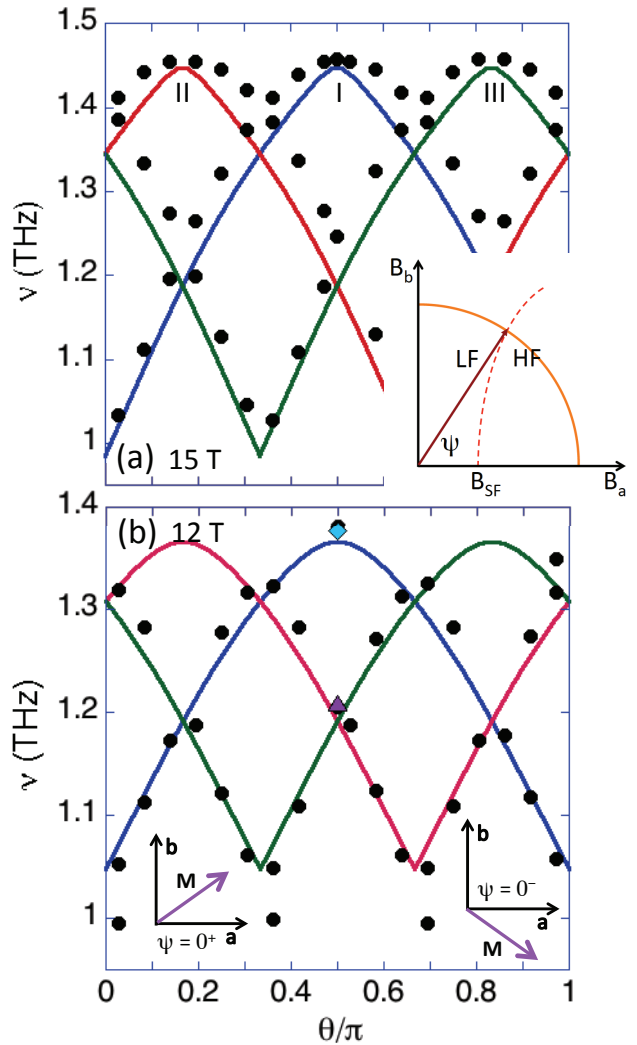}
\caption{(Color online) The measured (solid circles) and predicted (blue, red, and green curves for domains I, II, and III, respectively)
angular dependence of the MR mode frequencies for 12 and 15 T.  The inset to (a) sketches the angular dependence of $\vB_{\rm SF}$ (dashed curve), which separates
low-field (LF) and high-field (HF) states.  The inset to (b) shows the net magnetization of any domain for angles $\psi $ on either side of 0.  
Flips of the $b$-axis spin at $\psi =0$ and $\pi $ produce cusps in the mode frequencies.}
\end{figure}

\section{Fitting results}

Fits for the coupling parameters utilize the field dependence of $\vM $, the zero-field powder-diffraction data [\onlinecite{caig10}], the field dependence of the 
MR and EM modes at $\theta = \pi/2$ [\onlinecite{bord15}], and the MR mode frequencies at $\theta =0 $ and $\pi /3$ for 7, 12 and 15 T.
The resulting exchange and anisotropy constants are provided in Table I and
the corresponding zero-field spin state is plotted in Figs.~1(a-b).   In contrast to the previously proposed [\onlinecite{caig10}] spin state with zig-zag chains in the 
$ab$ plane containing spins 2, 3, 6, and 7, our spin state can be better described as an array of $c$-axis chains or 
connected bitetrahedra [\onlinecite{valldor04, chap06}] containing spins $\{1,2,3,4\}$ (chain $\alpha $) or $\{5,6,7,8\}$ (chain $\beta $) as sketched in Fig.~2.  
Chains are coupled by exchanges $J_9$, $J_{10}$, and $J_{12}$ in the $ab$ plane.

What explains the wide range of $J_i$ values?  An orthorhombic distortion [\onlinecite{caig10, panja16}] with $b/(a\sqrt{3})-1 \rightarrow 0.018$ as $T\rightarrow 0$
breaks the hexagonal symmetry of the $ab$ plane and explains the difference between the pairs $\{ J_1 ,J_2\}$, 
$\{ J_8, J_{11}\}$, and $\{J_{10}, J_{12}\}$.
The difference between couplings like $\{J_7 ,J_8\}$ is caused by charge ordering:  whereas $J_7$ couples moments
2 and 3 with $M_2=M_3$, $J_8$ couples moments 2 and 4 with $M_2\ne M_4$.  Charge ordering also explains the difference between the pairs
$\{J_2, J_3\}$ and $\{J_9 ,J_{10}\}$.  Although not demanded by symmetry, we set $J_1=J_5$, $J_2=J_4$, $J_3=J_6$, and $J_{11}=J_{12}$ 
because the spin state and excitations at $\vQ $ only depend on their averages [\onlinecite{average}].   

\begin{table*}
\caption{Ratios of powder-diffraction peak intensities}
\begin{ruledtabular}
\begin{tabular}{ccccccccc}
& $I(002)/I(101)$ & $I(012)/I(101)$ & $I(111)/I(101)$ & $I(112)/I(101)$ & $I(121)/I(101)$ & $I(122)/I(101)$ & $M_b (\mu_{\rm B}$/f.u.) & $\chi^2 $   
  \\
 \hline    
 experimental [\onlinecite{caig10}]  & 0.344 & 0.326 & 0.322 & 0.477 & 0.262 & 0.404 & &  \\                        
 previous [\onlinecite{caig10}]  & 0.286 & 0.414 & 0.384 & 0.411 & 0.232 & 0.427 & 0.88 & 0.021 \\
 currrent & 0.360 & 0.380 & 0.378 & 0.286 & 0.313 & 0.449 & 1.33 & 0.047 \\
\end{tabular}
\end{ruledtabular}
\label{alf}
\end{table*}

Given other conditions, our fit chooses the spin state that matches the powder-diffraction data [\onlinecite{caig10}] as closely as possible.
At zero field, the predicted spin state has angles $\phi_1= -0.83 \pi $, $\phi_2= 0.40\pi$, $\phi_3= -0.23\pi $, and $\phi_4= 0.62\pi $.  
Based exclusively on powder-diffraction data and symmetry constraints, the previously proposed
spin state [\onlinecite{caig10}] had $\phi_1=-0.24 \pi $, $\phi_2 = \phi_3 =0.67\pi $, and $\phi_4=-0.44\pi $.
In both cases, $\phi_{i+4} = \pi -\phi_i$ ($i=1,\ldots, 4$) so that the moment $M_b$ lies along the $b$ axis.
As shown in Table II, our spin state does not satisfy the powder diffraction data quite as well as the earlier state,
primarily because it underestimates the powder diffraction peak $I(112)$.

For the previous spin state, $\chi^2$ is minimized by Lorentzian form factors with $Q_1/4\pi = 0.088$ \AA$^{-1}$, 
$Q_2/4\pi = Q_3/4\pi = 0.095$ \AA$^{-1}$, and $Q_4/4\pi = 0.088$ \AA$^{-1}$ for spins $S_n$.
For the new spin state, $Q_1/4\pi = 0.052$ \AA$^{-1}$, $Q_2/4\pi = Q_3/4\pi = 0.224$ \AA$^{-1}$, and $Q_4/4\pi = 0.102$ \AA$^{-1}$.
All are smaller than the scale $Q_0/4\pi \approx 0.3 \,\AA^{-1}$ measured by Khan and Erickson [\onlinecite{khan70}] for Co$^{2+}$ in CoO.

Our results indicate that the exchange coupling $J_8\approx 188$ meV
between moments 2 (Co$^{2+}$, $S_2 = 1$) and 4 (Co$^{3+}$/Co$^{2+}\underline{L}$, $S_4 = 1.2$) is strongly ferromagnetic and 
larger in magnitude even than the 155 meV antiferromagnetic coupling found in the cuprate Nd$_2$CuO$_4$ [\onlinecite{bourges97}].
The strength of this coupling might be explained by the double-exchange mediated hopping of 
ligand holes $\underline{L}$  [\onlinecite{maig06}] from site 4 to 2.  Bear in mind, however, that
the estimated exchange parameters would be significantly reduced if the fits were performed with $S=3/2$ for all Co spins.   
In particular, $J_8$ would then fall from 188 to 100 meV.

Except for $J_{10}$, the five largest exchange couplings $J_1= J_5\approx -92$ meV, $J_3=J_6\approx 41$ meV, and $J_8\approx 188$ meV lie within
connected bitetrahedral, $c$-axis chains.  Inside each chain, competing interactions between spins 1, 2, and 4 produce
a non-collinear spin state.

Although occupying a triangular lattice, chains $\alpha $ 
and $\beta $ are magnetically ordered with
moments $\vM^{\rm ch}= (\pm 1.18,1.33,0)\mb $/f.u..  These chains are primarily coupled by 
the strongly ferromagnetic interaction $J_{10} \approx 108$ meV between nearly parallel 
spins $\{4, 6\}$ ($\phi_4=0.62\pi$, $\phi_6=0.60\pi $) and $\{2,8\}$ ($\phi_2 = 0.40\pi $, $\phi_8=0.38\pi $).
Above $T_c=70$ K, the short-range order within each chain may 
be responsible for the large, negative Curie-Weiss temperature  
$\Theta_{\rm CW} \approx -1720$ [\onlinecite{caig09}] or $-890$ K [\onlinecite{qu11}], the larger than 
expected Curie constant [\onlinecite{caig09}], and the susceptibility anomaly [\onlinecite{qu11}] at 360 K suggestive of 
short-range magnetic order far above $T_c$.

Comparison between the theoretical and experimental results for the magnetization in Fig.~3 suggests that roughly 20\% of the sample is in domain I.
Different domain populations or even orthorhombic twinning in other samples may 
explain the discrepancies between the reported magnetization measurements [\onlinecite{caig09, caig10, qu11, iwamoto12, seikh12}].

Easy-axis anisotropies $A$ and $C$ favor ferrimagnetic alignment along $\vb $ rather than $\va $.  The spin-flop (SF) 
field required to flip the spins towards the $\va $ direction must 
increase as the field along $\vb $ increases [\onlinecite{trans}].  As shown in the inset to Fig.~5(a), $B_{\rm SF}(\psi )$ then increases with $\psi $.
If $B_{\rm SF}(\psi = 0)  < 15$ T, then the MR spectrum
for 15 T would show a discontinuity at the transition from a low-field (LF) to a high-field (HF) state below some critical value of $\psi $.  
Since the MR mode frequencies in Fig.~5(a) do not exhibit any discontinuities as a function of $\psi $, 
we conclude that $B_{\rm SF}(\psi = 0)$ exceeds 15 T 
and probably, based on the smooth dependence of the magnetizations on field,
exceeds 32 T as well.  The apparent small size of $B_{\rm SF}$ [\onlinecite{pralong11, caig13}] must reflect the 
net magnetization of all three domains.

Predicted modes below 5 THz are plotted in Fig.~4.  The Goldstone modes for all three domains are lifted by in-plane anisotropies to 
become the MR modes with zero-field frequencies of 1.07 THz.   As remarked earlier, the lower MR mode comes from domains II and III while the 
upper MR mode comes from domain I.  
Below 3.5 THz, one EM mode is produced in domain I and another in domains II and III.   The degenerate EM modes from domains II and III dominate the
optical absorption.  The predicted field dependence of the upper MR mode is quite close to the observed dependence.
But the predicted curvatures of the lower MR mode and the EM mode, both from domains II and III, is not observed.  
  
 \section{Spin-Induced Electric Polarization} 
 
Below the ferrimagnetic transition, \CBO is reported [\onlinecite{caig13}] to develop a very large spin-induced 
polarization $\sim 1700$ nC/cm$^2$, which is surpassed in type I multiferroics only by the conjectured  [\onlinecite{jun15}] 
spin-induced polarization $\sim 3000$ nC/cm$^2$ of BiFeO$_3$.  Other measurements indicate that the spin-induced polarization of \CBO 
ranges from 320 nC/cm$^2$ [\onlinecite{iwamoto12}] to 900 nC/cm$^2$ [\onlinecite{kocsis}].   
  
The electric-field dependence of any interaction term in the spin Hamiltonian ${\cal H}$ can induce an electric polarization below $T_c$.   
However, the electric-field dependence of the easy-plane anisotropy $D$ cannot explain the 
spin-induced polarization along $\vc $ because the expectation value of $P_i = \kappa {S_{ic}}^2$ with $\kappa = -\partial D/\partial E_c$
would vanish in zero magnetic field when all the spins lie in the $ab$ plane.  Easy-axis anisotropy $A$ or $C$ in the $ab$ plane could
produce a spin-induced electric polarization perpendicular to $\vc $.  But the EM mode would then become observable for a THz electric field in the $ab$ plane,
contrary to measurements.

As conjectured previously [\onlinecite{caig13}], 
the spin-induced polarization in \CBO must then be generated by the dependence of the exchange interactions $J_{ij}$ on an electric field, called
magnetostriction.  Coupling constant $\lambda_{ij}= \partial J_{ij} /\partial E_c$ for bond $\{ i,j \}$ is associated with 
a spin-induced polarization [\onlinecite{PP}] per site of $P_c^{ij} = \lambda_{ij}\, \vS_i\cdot \vS_j /4$, which accounts for the four equivalent bonds per unit cell.
Expanding in the electric field $E_c$ yields an interaction term $-E_c\, \lambda_{ij}\, \vS_i \cdot \vS_j$, linear in the electric field and quadratic in the spin
operators.

Taking $\vert 0\rangle $ as the ground state and $\vert n\rangle $ as the excited spin-wave state,
the MR matrix element $\langle n \vert M_a \vert 0\rangle $ mixes with the EM matrix element
$\langle n \vert P^{ij}_c \vert 0\rangle $ for domains II and III but not for domain I. 
Therefore, our model can explain the strong asymmetry [\onlinecite{miyahara11}] $\sim {\rm Re} \bigl\{ \langle n \vert \vM\cdot \vB_{\omega}  \vert 0\rangle 
\langle 0 \vert \vP \cdot \vE_{\omega} \vert n\rangle \bigr\}$ in the absorption of counter-propagating light waves [\onlinecite{bord15}] for the lower observed MR mode
in geometry $ii$ with $\vE_{\omega } || \vc $.  But it cannot explain the observed asymmetry of this mode in geometry $i$ with ${\bf E}_{\omega } \perp \vc $
if only $\langle 0 \vert P_c\vert n \rangle $ is significant.  

How can we estimate the coupling constants $\lambda_{ij}$ and the spin-induced electric polarization?  
The optical absorption of any mode in domain $l$ is proportional to ${p_l}^2$ because each matrix element is separately proportional to $p_l$.
At nonzero field, the EM mode absorption is proportional to ${p_2}^2+{p_3}^2$ while the upper MR mode absorption is proportional to ${p_1}^2$. 
At zero field, all domains have the same mode spectrum so that both the MR and EM mode absorptions are proportional to ${p_1}^2+{p_2}^2+{p_3}^2$. 
Experimentally, the ratio $r$ of the absorption of the EM mode to the absorption of the upper MR mode rises from $r=7.5$ at 0 T to $r=35$ at 10 T.  
This growth is explained by the $B > 0$ ratio $({p_2}^2+{p_3}^2)/{p_1}^2 \approx 10 \pm 5$ for $p_1=0.185\pm 0.071$ and $p_2=p_3=(1-p_1)/2$.

At both 0 and 10 T, the only sets of bonds that generate spin-induced polarizations of the right magnitude are $\{2,7\}$ and $\{3,4\}$.
Each of those bonds couples adjacent $c$-axis chains through pairs of spins that are almost anti-parallel.
From the relative absorptions $r$ at 0 or 10 T, we estimate that $\langle P^{27}_c \rangle \approx 2350$ or 1960 nC/cm$^2$ 
and $\langle P^{34}_c \rangle \approx 2110$ or 1730 nC/cm$^2$.
Results for both sets of bonds are consistent with the recently observed [\onlinecite{caig13}] polarization of 1700 nC/cm$^2$.
By contrast, density-functional theory [\onlinecite{john14}] predicts that the spin-induced polarization along $\vc $ is 460 nC/cm$^2$. 
The spin-induced polarization should remain fairly constant with applied magnetic field, decreasing by about 1\% for a 10 T field along $\vb $.

\section{Conclusion}

We have presented a nearly complete solution for the magnetization, spin state, and mode frequencies of the swedenborgite \CBP.  
An orthorhombic distortion above $T_c$ partially relieves the geometric frustration on the kagome and triangular layers
and allows ferrimagnetism and ferroelectricity to coexist below $T_c$.
Although occupying a triangular lattice, bitetrahedral $c$-axis chains are ferrimagnetically ordered in the $ab$ plane.  Competing interactions 
within each chain produce non-collinear spin states.  Sets of bonds coupling those chains are responsible for the 
large spin-induced polarization of \CBP.

Despite its fixed permanent electric polarization, this swedenborgite may yet have important technological applications 
utilizing the large changes [\onlinecite{caig13}] in the spin-induced
polarization when a modest magnetic field $< 1$ T is applied along $\vb $ just below $T_c$.  
A big jump in the polarization should also be produced just below $T_c$ by rotating a fixed magnetic field about the $c$ axis.  
Above all, our work illuminates a pathway to develop other functional materials with sizeable magnetic moments and electrical polarizations. 

\acknowledgements

Research sponsored by the U.S. Department of Energy, Office of Science, Basic Energy Sciences, Materials Sciences and Engineering Division (RF),
by the Hungarian Research Funds OTKA K 108918, OTKA PD 111756, and Bolyai 00565/14/11 (SB, VK, and IK), and
by the  institutional research funding IUT23-3 of the Estonian Ministry of Education and Research
and the European Regional Development Fund project TK134 (TR and UN).


\begin{thebibliography}{}
\bibitem{experiment}  The THz spectra of \CBO were measured at 2.5 K with linear light polarization using Fourier-transform spectroscopy [\onlinecite{bord15}].
The reference spectrum was taken in zero field.
\bibitem{average} Spin excitations away from $\vq = \vQ $ depend separately on these exchange constants.  The stability of the 
coplanar ground state requires that each pair of exchange constants be sufficiently close to one another.
\bibitem{trans}  While the coplanar LF spin state has a lower energy than the HF state below 32 T, it becomes locally unstable to 
a buckled state with spins canted out of the $ab$ plane between 16 and 32 T for field along $\va $.  Based on the smooth dependence of the
magnetizations on field, the transition from the coplanar to the buckled state must be second order.
\bibitem{PP} Based on the symmetry of adjacent layers, it is easy to show that the polarization operator 
$\vP^{12}$ associated with bond $\{1,2\}$ has components
\begin{eqnarray}
P_a^{12} & \propto & C_{12}- C_{56} -C_{1^{\prime } 2^{\prime }} + C_{5^{\prime }6^{\prime }}, \nonumber \\
P_b^{12} & \propto & C_{12} +C_{56} -C_{1^{\prime } 2^{\prime }} - C_{5^{\prime }6^{\prime }}, \nonumber \\
P_c^{12} & \propto & C_{12}+ C_{56} +C_{1^{\prime } 2^{\prime }} + C_{5^{\prime }6^{\prime }}, \nonumber 
\end{eqnarray}
where $C_{ij}= \vS_i\cdot \vS_j $.  Similar relations hold for other bonds.  Only the $c$ component $P_c^{ij}$ can produce a static polarization
but all three components may contribute to the off-diagonal polarization matrix elements $\langle 0\vert P_{\alpha }^{ij} \vert n \rangle $ ($n\ne 0$).
\end{thebibliography}
\end{document}